\documentclass[12pt]{article}
\pdfoutput=1

\usepackage[DIV13]{typearea}
\usepackage{graphicx}
\usepackage{amsmath}
\usepackage{amsfonts}
\usepackage{mathrsfs}
\usepackage{amssymb}
\usepackage{subfigure}
\usepackage{multicol}
\usepackage[usenames,dvipsnames]{color}
\usepackage{feynmp}
\usepackage{multirow}
\usepackage{array}
\usepackage{slashed}
\usepackage{units}
\usepackage{bbm}
\usepackage[normalem]{ulem}
\usepackage{framed}
\usepackage[left=.9in, right=.9in]{geometry}
\usepackage[titletoc,title]{appendix}
\newcolumntype{L}[1]{>{\raggedright\let\newline\\\arraybackslash\hspace{0pt}}m{#1}}
\newcolumntype{C}[1]{>{\centering\let\newline\\\arraybackslash\hspace{0pt}}m{#1}}
\newcolumntype{R}[1]{>{\raggedleft\let\newline\\\arraybackslash\hspace{0pt}}m{#1}}

\usepackage{hyperref}
\hypersetup{
 colorlinks,
 citecolor=Red,
 linkcolor=Blue,
 urlcolor=Blue}

\begin{document}

\vspace{0.5cm}

\begin{center}
{\Large\bf Theoretical Constraints on\\[0.1in] Neutron--Mirror-Neutron Oscillation}
\end{center}

\vspace{0.5cm}
\begin{center}
{\large \bf K.S. Babu}$^{a,}$\footnote{E-mail: babu@okstate.edu} {\large and}
{\large \bf Rabindra N. Mohapatra}$^{b,}$\footnote{E-mail: rmohapat@umd.edu}

\vspace{0.5cm}

\centerline{$^{a}${\it  Department of Physics, Oklahoma State University, Stillwater, OK 74078, USA }}

\vspace*{0.2cm}
\centerline{$^{b}${\it Maryland Center for Fundamental Physics, Department of Physics, University of Maryland,}} 
\centerline{\it College Park, MD 20742, USA}

\end{center}
\vspace{0.6cm}

\begin{abstract}

Mirror models lead to the possibility that neutron ($n$) can oscillate into its mirror partner ($n'$) inspiring several experimental searches for this phenomenon. The condition for observability of this oscillation is a high degree of degeneracy between the $n$ and $n'$ masses, which can be guaranteed if there is exact parity symmetry taking all particles to their mirror partners. However consistency of these models with big bang nucleosynthesis requires that this parity symmetry be broken in the early universe in a scenario called asymmetric inflation. In this paper we study the consistency of an observable $n-n'$ oscillations signal with asymmetric inflation and derive various theoretical constraints. In particular, we find that the reheat temperature after inflation should lie below 2.5 TeV, and predict a singlet fermion with a mass below 100 GeV. In simple models where the right-handed neutrino is a mediator of baryon number violating interactions we find that the light neutrinos are Dirac fermions with their masses arising radiatively through one-loop diagrams. 
\end{abstract}

\section{Introduction} 

The mirror models~\cite{mirror} were proposed  many years ago by Lee and Yang as a possible way to reconcile the observed parity violation in weak interactions with a more fundamental theory that conserves parity. They proposed that parity could be taking observed particles to their parity partners in a different  world which interacts with our world only via gravity. The second world is called the mirror world in recent literature.  The success of the standard model (SM) has provided a very specific platform for exploring more detailed phenomenological implications of this idea. In this framework one duplicates all the SM particles and  forces in the mirror sector with identical couplings.. While both sides were initially assumed to be connected only by gravitational interactions as stated, they could also be connected by other fields that are SM singlets. This basic setup  has the implication that there is a duplication of all observed SM baryons and leptons in the mirror sector, which are dark particles as far as the visible world is concerned.  Mirror models could therefore provide candidates for dark matter  of the universe with mirror baryons playing that role. This framework also has the potential to solve the problems of baryogenesis, neutrino masses and  inflation by extending the particle spectrum of each sector in a symmetric way. Thus in principle mirror models could provide a unified platform for solving many of the puzzles of the standard model.

There are two distinct realizations of the mirror models in the literature: one class, called symmetric mirror model has all scales in the mirror sector the same as the corresponding scales in the visible sector \cite{mirror2}, and a second class where the two weak scales are different after spontaneous parity breaking. In the former class of models, it has been pointed out that if there are new interactions beyond the standard model connecting the two sectors, there will occur a new phenomenon known as neutron-mirror neutron ($n-n'$) oscillation~\cite{nnprime}. Since our knowledge about the mirror sector of the universe is very limited, such oscillations can take place with low oscillation time (or high rate) without conflicting with any other observations. There are currently several experiments searching for this process~\cite{expts}. 

The evolution of a mixed $n-n'$ system in vacuum and in the absence of magnetic fields is governed by the evolution equation
\begin{eqnarray}
\frac{d}{dt}\left(\begin{matrix} n \cr n' \end{matrix} \right) = \left(  \begin{matrix} m & \delta \cr \delta & m'  \end{matrix}\right)\left(\begin{matrix} n \cr n' \end{matrix} \right)~.
\label{Ham}
\end{eqnarray}
With initial beam of neutrons the probability for the appearance of mirror neutron $n'$ at time $t$ is given by
\begin{eqnarray}
P_{n-n'}= \frac{4\delta^2}{(\Delta ^2+4\delta^2)} \sin^2\frac{\sqrt{\Delta^2+4\delta^2}\,t}{2}
\end{eqnarray}
where $\Delta = m_n-m_{n'}$ and $\delta$ is the $n-n'$ mixing parameter in the Lagrangian $\delta (n n')$.  This coupling, which violates baryon number $B$, but preserves $(B-B')$  where $B'$ is the mirror baryon number, should arise from some new interactions in  the quark Lagrangian that connects the two sectors. Since the typical neutron transit time is of order 1 second, in order for $n-n'$ oscillation to be observable, $\Delta \leq 10^{-24}$ GeV is required. For larger values of $\Delta$, the oscillation probability is suppressed as
$P_{n-n'}
\simeq \left(\frac{\delta}{\Delta}\right)^2$, which becomes unobservable. For a discussion of $n-n'$ oscillation in presence of magnetic field and matter see Ref. \cite{KB}. 
It is therefore important to estimate how large $\Delta$ can be in different realistic theoretical frameworks for BSM physics that lead to observable  $n-n'$ oscillation, while implementing asymmetric inflation so as to be consistent with big bang nucleosynthesis (BBN).

Asymmetric inflation was proposed in the context of mirror models~\cite{BDM} as a way to reconcile the success of BBN with the extra relativistic degrees of freedom present in the mirror model. In the standard BBN theory  during the epoch of nucleosynthesis the light degrees of freedom present in the plasma are three light neutrinos, electron, the photon and their antiparticles, leading to an effective degree of freedom of $g^* = 10.75$. This is is consistent with light element abundances derived from BBN.  In the mirror model with symmetric inflation the effective number of light degrees would be doubled to $g^* = 21.5$ during BBN, which would spoil its success.  Asymmetric inflation reduces this abundance by realizing the temperature of the mirror world to be a factor of 1/3 or so below that of the SM plasma.  Since energy density scales as the fourth power of $T$, the effective number of degrees would now be lowered by a factor of $(1/3)^4 \simeq 0.012$. The effective light degrees from the mirror world would be then decreased to 0.13, which is consistent with BBN. Other ways of reducing the number of light degrees, such as by late decay of particles dumping entropy, would appear fortuitous in the mirror framework due to the large number of extra degrees present in the theory.

Neutron-mirror neutron oscillation is similar in many respects to neutron-anti-neutron ($n-\overline{n}$)  oscillation~\cite{nnbar}. The main difference between the two are two-fold: (i) The degeneracy of $n$ and $\bar{n}$ required for oscillation to take place is guaranteed by CPT invariance which holds in all local and Lorentz invariant theories. On the other hand, the mass degeneracy  between $n$ and $n'$ requires that there be an exact parity symmetry that transforms the visible constituents and forces in the SM to the mirror sector. However, in order for these models to be consistent with the success of BBN, the universe must undergo a phase of  asymmetric inflation at very early moments which
makes the reheat temperature of the mirror sector less than that of the visible sector~\cite{BDM}.  This introduces asymmetry in the spectrum of particle masses between the two sectors.~\cite{MN}. In particular, this will break the mass degeneracy required for efficient $n-n'$ oscillation.  (ii) The mirror neutron does not feel the normal nuclear force. As a result, a bound neutron cannot undergo $n-n'$ oscillation except in the neutron star where the binding force is gravity. However a bound neutron can undergo $n-\bar{n}$ oscillation. The only bound environment where $n-n'$ oscillation can take place is inside the neutron star where  the main binding force is gravity. For constraints on $n-n'$ oscillation arising from pulsars and neutron stars see Ref. \cite{RN}. 

Currently the strongest limit on $n-n'$ oscillation time, defined as $\tau_{nn'} = \hbar/\delta$ is $\tau_{nn'} \geq 352$ sec. \cite{expts}, which implies $\delta \leq 2 \times 10^{-27}$ GeV. Ongoing and proposed experiments are expected to improve this limit considerably.  These searches are also motivated by one interpretation of the neutron lifetime puzzle in terms of $n-n'$ oscillations \cite{bere3} which prefers a range for $m-m' \approx 10^{-18}$ GeV and $\delta/\Delta \approx 10^{-2}$.  Such values will be shown to arise naturally in the models we develop here.

Our main goal in this paper is to study the consistency between asymmetric inflation and an observable $n-n'$ oscillation in  realistic mirror models which also accommodate non-vanishing neutrino masses. We find it to be quite constraining. There are several sources for these constraints.  First, the two sectors should not be in equilibrium after inflationary reheating in order to be consistent with BBN.  This, when combined with observable $n-n'$ oscillation signal sets an upper limit on the reheat temperature of 2.5 TeV.  We find that the effective $B$-violating interaction can be generated in renormalizable models in a unique way, which requires a singlet fermion with a mass below 100 GeV.  When this fermion is identified as the right-handed neutrino which takes part in neutrino mass generation, two possible scenarios arise: one where the neutrinos are Majorana fermions which get their small mass via the seesaw mechanism, with one flavor decoupled from the rest; and a second, more symmetric one where the neutrinos are Dirac fermions. In the second scenario we show that small Dirac masses arise as radiative corrections through one-loop diagrams, somewhat similar to the case in scotogenic models.  In both scenarios it is possible to have observable $n-n'$ as well as $n-\overline{n}$ oscillations, which are accessible to ongoing and forthcoming experiments.

The rest of the paper is organized as follows. In Sec. 2 we describe a minimal mirror model framework which is compatible with asymmetric inflation.  Here we show why the $n-n'$ mass splitting makes oscillations highly suppressed.  Various constraints arising from the two sectors not establishing thermal equilibrium are also derived here.  In Sec. 3 we present a modified framework which has $B$ violating interactions. Here we derive an upper limit on the reheat temperature. We also estimate $n-n'$ mass splitting and show that there exist parameters which are consistent with observable oscillations. In Sec. 4 we draw a connection between $n-n'$ oscillations and neutrino mass generation. Here we develop a Dirac neutrino mass model where the masses arise as one-loop radiative corrections. Finally in Sec. 5 we conclude.


\section{A minimal mirror model framework}
\label{sec2}

We begin with a description of a minimal framework where the mirror parity symmetry can be realized  and  asymmetric inflation can be successfully implemented.  As we shall see, this minimal framework will not lead to observable $n-n'$ oscillations for two reasons: Baryon number is separately conserved in the SM sector and in the mirror sector (denoted as SM$'$); and the mass splitting between $n$ and $n'$ turns out to be too large when asymmetric inflation is implemented, which suppresses $n-n'$ oscillations well below current experimental sensitivity. Nevertheless, this framework would serve as our starting point for a realistic scenario where $n-n'$ oscillations can be in the observable range. This  framework would also provide insight into the modifications needed  in order to circumvent various theoretical constraints to bring  $n-n'$ oscillations into the observable range.

The gauge symmetry of the minimal mirror model is $[SU(3)_c \times SU(2)_L \times U(1)_Y] \times [SU(3)'_c \times SU(2)'_L \times U(1)'_Y]$.  The particle content of the mirror sector (indicated with a prime) is identical to that of the SM sector, but transforming under the mirror gauge symmetry. 
The Lagrangian of the model is the SM Lagrangian and its mirror replica, as well as new interaction terms needed to implement asymmetric inflation through a parity odd singlet scalar field $\eta$.  It is given by
\begin{equation}
    {\cal L} = {\cal L}_{\rm SM} + {\cal L}_{\rm SM'} + {\cal L}_{\rm new}
    \label{min}
\end{equation}
where under the mirror parity symmetry $ {\cal L}_{\rm SM} \leftrightarrow {\cal L}_{\rm SM'}$ and ${\cal L}_{\rm new}$ is mirror parity invariant (see below).  This implies that the parameters of ${\cal L}_{\rm SM'}$ are identical to those of ${\cal L}_{\rm SM}$, which makes the scenario very predictive. We define ${\cal L}_{\rm SM}$ to contain right-handed neutrino fields $N_R$, which are used for generating small neutrino masses via the seesaw mechanism. Similarly ${\cal L}_{\rm SM'}$ contains the $N_R'$ fields that generate small mirror neutrino masses. Mirror parity implies that the masses and couplings of the $N$ and $N'$ fields are identical. Specifically, the seesaw sector of the Lagrangian contains the terms (along with the kinetic energies of $N$ and $N'$)
\begin{equation}
    {\cal L}_{\rm seesaw} = - \frac{1}{2}\left(\overline{N_R^c} M_N^0 N_R + \overline{N_R^{\prime c}} M_N^0 N'_R\right) + \left(\overline{L}Y_D \tilde{H} N_R + \overline{L'}Y_D \tilde{H'} N'_R \right) + h.c.
    \label{seesaw}
\end{equation}
where $L^T = (\ell,~ \nu)$ denotes the left-handed lepton doublets and $\tilde{H} = i \tau_2 H^*$ with $H$ being the Higgs doublet with $(Y/2) = 1/2$, along with similar definitions for the primed fields.

${\cal L}_{\rm new}$ of Eq. (\ref{min}) contains interactions terms involving  a parity odd inflaton field $\eta$ which is  a real scalar singlet under the SM and SM$'$ gauge symmetries.  It also contains cross terms between the two sectors, and is given by
\begin{eqnarray}
-{\cal L}_{\rm new} &=& -\frac{1}{2}(\partial^\mu \eta)\partial_\mu \eta) +\frac{\chi}{2} B_{\mu \nu} B^{\prime \mu\nu}+ \frac{m_\eta^2}{2} \eta^2 + \frac{\lambda_\eta}{4!}\eta^4 + \mu_\eta \eta(H^\dagger H - H^{'\dagger} H') \nonumber \\
&+& \lambda_{\eta H} \eta^2(H^\dagger H + H^{'\dagger} H') + \frac{\eta}{2} \left\{ (\overline{N_R^c} Y_\eta N_R - \overline{N_R^{'c}} Y_\eta N'_R) + h.c. \right\} \nonumber \\
&+& \lambda_{HH'} (H^\dagger H)(H^{'\dagger} H')~.
\label{new}
\end{eqnarray}
We now turn to the mechanism of realizing asymmetric inflation from this Lagrangian and derive  constraints on the model parameters for achieving this consistently.

\subsection{Realizing asymmetric inflation}

The scalar field $\eta$ acquires a vacuum expectation value (VEV) $\left\langle \eta \right\rangle = v_\eta$ which breaks the mirror parity symmetry spontaneously.  This is achieved by choosing $\mu_\eta^2 < 0$ in Eq. (\ref{new}). The $\eta$ field is also the inflaton, with its potential sufficiently flat so that the slow roll conditions are satisfied.  With only the renormalizable quadratic and quartic terms for $\eta$ in the potential, as in Eq. (\ref{new}), chaotic  inflation is realized, which is generally successful.  However, it turns out that detailed predictions for the spectral index $n_s$ and the tensor to scalar ratio $r$ lie slightly outside the range allowed by observations by the Planck satellite \cite{Planck}.  This conflict can be overcome by non-minimally coupling the $\eta$ field to gravity, see for example~\cite{BS,OSS}.  This is the scenario of inflation that we shall adopt here. (For a recent analysis showing the  consistency of this approach see Ref. \cite{shafi1}.)

In the standard picture of inflationary reheating, the inflaton decays while it oscillates around the minimum of its potential.  The energy stored in the inflaton field is transferred into its relativistic decay products in the process. These daughter particles thermalize and constitute a radiation dominated universe.  The reheat temperature is determined by the decay width of the inflaton $\eta$ into daughter particles.  The decay width $\Gamma(\eta \rightarrow {\rm SM})$ differs from the width  $\Gamma(\eta \rightarrow {\rm SM'})$, since mirror parity symmetry is broken, leading to an asymmetry in the reheat temperatures.  This asymmetry is essential for the mirror model to be compatible with big bang nucleosynthesis. We now show how this can happen in the minimal mirror model framework given in Eqs. (\ref{min})-(\ref{new}).

Once $\eta$ field develops a VEV, the masses of the Higgs doublet $H$ and its mirror partner $H'$ split and are given by
\begin{eqnarray}
M_H^2 &=& \mu_H^2 + \lambda_{\eta H} v_\eta^2 + \mu_\eta v_\eta \nonumber \\
M_{H'}^2 &=& \mu_H^2 + \lambda_{\eta H} v_\eta^2 - \mu_\eta v_\eta~
\end{eqnarray}
where $\mu_H^2$ is the coefficient of the $(H^\dagger H)$ term in the SM Lagrangian. The widths for the decay of $\eta$ into $H^\dagger H$ and $H^{\prime \dagger}H'$ are given by
\begin{eqnarray}
\Gamma(\eta \rightarrow H^\dagger H) &=& \frac{(2 \lambda_{\eta H} v_\eta + \mu_\eta)^2}{8 \pi M_\eta} \sqrt{1-4 \frac{M_H^2}{M_\eta^2}} \nonumber \\
\Gamma(\eta \rightarrow H^{\prime \dagger} H') &=& \frac{(2 \lambda_{\eta H} v_\eta - \mu_\eta)^2}{8 \pi M_\eta} \sqrt{1-4 \frac{M_{H'}^2}{M_\eta^2}}~.
\end{eqnarray}
In the radiation dominated era, the Hubble expansion rate is given by
\begin{equation}
    H(T) = \left[{\frac{4}{45}\pi^3 g^*(T)}  \right]^{1/2} \frac{T^2}{M_P} = 1.66 \sqrt{g^*(T)} \frac{T^2}{M_P},
\end{equation}
where $g^*(T)$ is the effective light degrees of freedom in equilibrium with the plasma at temperature $T$ and $M_P = 1.22 \times 10^{19}$ GeV is the Planck mass. The reheat temperature $T_{RH}$ is obtained by equating the Hubble rate to the decay rate of the inflaton into the SM fields -- the Higgs field in the present case. This yields
\begin{equation}
    T_{RH} = \sqrt{\frac{M_P\Gamma(\eta \rightarrow H^\dagger H)}{1.66 \sqrt{g^*(T_{RH})}}}~.
\end{equation}
Similarly, the reheat temperature $T_{RH}'$ of the mirror world is obtained by equating the decay rate of $\eta$ into the mirror sector.  We therefore obtain, in the approximation $M_\eta^2 \gg M_H^2$, a relation
\begin{equation}
    \frac{T_{RH}'}{T_{RH}} =  \left|\frac{2 \lambda_{\eta H} v_\eta - \mu_\eta}{2 \lambda_{\eta H} v_\eta + \mu_\eta}\right|~.
\end{equation}
A desirable value of $(T'_{RH}/T_{RH}) \approx 1/3$, consistent with BBN constraints,  can be realized by choosing $\mu_\eta/(2 \lambda_{\eta H}v_\eta) \approx 1/2$. For such a choice, the mass parameter $\mu_\eta$ can be related to the reheat temperature  as
\begin{equation}
    \mu_\eta \approx (2.2 \times 10^{-3}~{\rm GeV}) \times \left[\frac{M_\eta}{10^8~{\rm GeV}}\right]^{1/2} \left[\frac{T_{RH}}{100~{\rm GeV}}\right]~,
\end{equation}
where we have set $g^*(T_{RH}) \approx 112$. 
We have normalized the reheat temperature to be relatively low here, since that will be shown to be a requirement for observable $n-n'$ oscillations in the next section.

One important constraint on $n-n'$ oscillation arises from this analysis where the inflaton decays asymmetrically to the Higgs fields of the two sectors.  The masses of the two Higgs fields are now split, with the splitting given by
\begin{eqnarray}
    M_H^2 - M_{H'}^2 = 2 \mu_\eta v_\eta 
    \approx \mu_\eta^2 \approx 5 \times 10^{-6} ~{\rm GeV}^2 \times \left[\frac{M_\eta}{10^8~{\rm GeV}}\right] \left[\frac{T_{RH}}{100~{\rm GeV}}\right]^2~.
\end{eqnarray}
Here in the second step we have shown the minimum value of the mass splitting, where  we used the relation $2 \lambda_{\eta H} v_\eta \approx 2 \mu_\eta$, and the fact that $\lambda_{\eta H}$ cannot be more than ${\cal O} (1)$.  This splitting would lead to a shift in the VEVs of $H$ and $H'$ fields, which for $M_\eta \sim 10^8$ GeV and $T_{RH}$ of order 100 GeV is of order
\begin{equation}
    \frac{v-v'}{v} \approx 10^{-6}~.
    \label{split}
\end{equation}
As a result the neutron and mirror neutron masses would split, with the splitting of order $m_n - m_{n'} \sim 10^{-8}$ GeV, which takes into account the fact that the Higgs VEV contributes to these masses only at the level of 1\%, with the dominant contribution arising from QCD and mirror QCD dynamics. This splitting is too large and would suppress $n-n'$ oscillations to a level unobservable in experiments.

A second way to realize asymmetric inflation in the same minimal framework is to utilize the coupling of $\eta$ with the right-handed neutrinos, as given in Eq. (\ref{new}).  Including the parity asymmetric contribution from $v_\eta$, the mass matrices for the $N$ and $N'$ fields are given by
\begin{eqnarray}
M_N &=& M_N^0 + Y_\eta v_\eta \nonumber \\
M_{N'} &=& M_N^0 - Y_\eta v_\eta~,
\end{eqnarray}
where $M_N^0$ is the common mass term for $N$ and $N'$ defined in Eq, (\ref{seesaw}).  
These mass matrices are diagonalized by unitary transformations $VM_N V^T = M_N^{\rm diag}$ and $V'M_{N'} V^{\prime T} = M_{N'}^{\rm diag}$.  The widths for $\eta$ decay into $N$ pairs and $N'$ pairs are given by
\begin{eqnarray}
\Gamma(\eta \rightarrow N N) &=& \sum_{i,j}\frac{ \left|(V^T Y_\eta V)_{ij}\right|^2}{8 \pi (1 + \delta_{ij})} M_\eta \left[1-\frac{(M_i + M_j)^2}{M_\eta^2}  \right]^{1/2} \left[ 1- \frac{(M_i - M_j)^2}{M_\eta^2}\right]~, \nonumber \\
\Gamma(\eta \rightarrow N' N') &=& \sum_{i,j}\frac{ \left|(V^{\prime T} Y_\eta V')_{ij}\right|^2}{8 \pi (1 + \delta_{ij})} M_\eta \left[1-\frac{(M'_i + M'_j)^2}{M_\eta^2}  \right]^{1/2} \left[ 1- \frac{(M'_i - M'_j)^2}{M_\eta^2}\right]
\label{kin}
\end{eqnarray}
where $M_i$ and $M'_i$ stand for the mass eigenvalues of the $N_i$ and $N'_i$ fields.  Unlike in the case of $\eta$ decays into Higgs fields, here an asymmetry would have to rely on the kinematic factors of Eq. (\ref{kin}), since in the limit of ignoring $M_i$ and $M'_i$ in relation to $M_\eta$, the decay rates become identical, in spite of the presence of two different unitary matrices $V$ and $V'$ in the decay rate formulas.  

\begin{figure}
    \centering
    \includegraphics[scale=0.55]{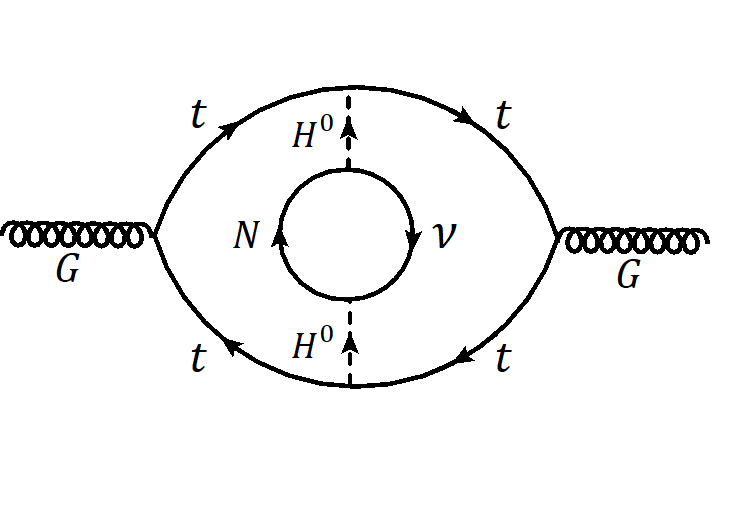}
    \caption{Feynman diagram that generates an asymmetry in the strong coupling $\alpha_s$ and its mirror counterpart $\alpha_s'$ in the minimal mirror framework where asymmetric inflation is realized through inflaton decays into $N$ and $N'$. The masses of $N$ and $N'$ are different in this case, leading to the asymmetry in the strong couplings.}
    \label{fig:fig1}
\end{figure}

If asymmetric inflation is realized via kinematics in these decays, there is still difficulty with observable $n-n'$ oscillations. To illustrate this constraint, consider $\eta$ decaying dominantly to one flavor of $N$ and $N'$ fields with masses $M_N$ and $M_N'$.  The ratio of reheat temperatures in the mirror sector to the SM sector is then given by
\begin{equation}
    \frac{T'_{RH}}{T_{RH}} = \left| \frac{M_\eta^2 - 4 M_N^{\prime 2}}{ M_\eta^2 - 4 M_N^2}\right|^{1/4}~.
\end{equation}
This ratio can be set to a value such as $1/3$ by a  suitable choice of $M_N$ and $M_N'$.  However, this would require $M_N$ and $M_N'$ to differ by at least order one.  Between these two mass scales the QCD coupling $\alpha_s$ and the mirror QCD coupling $\alpha_s'$ will evolve differently, as can be seen from a three loop diagram shown in Fig. 1.  Recall that the QCD scale parameter $\Lambda_{QCD}$ is defined through one-loop renormalization group evolution as
\begin{equation}
    \Lambda^2_{QCD} = \mu^2 e^{-\frac{4\pi}{\beta_0 \alpha_s(\mu^2)}},
\end{equation}
where $\beta_0 = 7$ is the one-loop beta function coefficient with six flavors of quarks.  An analogous expression can be written down for $\Lambda'_{QCD}$, the mirror QCD scale parameter.  In the momentum regime $M_N \leq \mu \leq M_N'$ the two couplings run differently owing to diagrams such as Fig. \ref{fig:fig1}. This results in a difference in the effective values of $\beta_0$ and $\beta_0'$ which we estimate  to be
\begin{equation}
    \beta_0 - \beta_0' \approx \frac{g_s^2 Y_t^2 Y_D^2}{(16 \pi^2)^2} {\rm ln}\left(\frac{M^2_{N^\prime }}{M_N^2} \right)~,
    \label{beta}
\end{equation}
where $Y_t$ is the top quark Yukawa coupling and $Y_D$ the Dirac neutrino Yukawa coupling defined in Eq. (\ref{seesaw}).  Now, the coupling $Y_D$ cannot be too small, as it is responsible to bring the $N$ ($N'$) states into equilibrium with the rest of the SM (SM$'$) plasma.  The width of $N$ decaying into $L + H$ and its CP conjugate states is given by
\begin{equation}
    \Gamma(N \rightarrow (L + H) +  (\overline{L} + H^*)) = \frac{|Y_D|^2}{8\pi} M_N~,
\end{equation}
which can compete with the Hubble rate at temperatures of order the $N$ mass  provided that the condition
\begin{equation}
    |Y_D| \geq 6 \times 10^{-8} \left[\frac{M_N}{100~{\rm GeV}} \right]^{1/2}
\end{equation}
is satisfied. Inserting this lower limit on $|Y_D|$ in Eq. (\ref{beta}), we obtain (for $M_N \approx 100$ GeV)
\begin{equation}
    \beta_0 - \beta_0' \approx 10^{-15}~.
\end{equation}
The difference causes a difference in the two QCD scale parameters which follows from Eq. (\ref{beta}):
\begin{equation}
    \frac{\Lambda_{QCD} - \Lambda'_{QCD}}{\Lambda} = \frac{2 \pi}{\alpha_s}\frac{\beta_0 - \beta_0'}{\beta_0^2} \approx 10^{-15}~.
\end{equation}
Consequently, the masses of $n$ and $n'$ will split by about $10^{-15}$ GeV -- as these masses arise primarily proportional to $\Lambda_{QCD}$ and $\Lambda'_{QCD}$, making $n-n'$ oscillations strongly suppressed.

To summarize, in this minimal mirror model framework, asymmetric inflation can be realized, but the mass splitting between the neutron and mirror neutron is too large for $n-n'$ oscillations to be observable. This framework also lacks baryon number violating interactions that are needed for $n-n'$ oscillations to occur.

\subsection{Constraints on mixed couplings from BBN}

The success of asymmetric inflation relies on the condition that the SM sector and the mirror sector do not come into equilibrium once reheating is established.  There are three sets of interaction terms in the minimal framework that can potentially bring the two sectors into thermal equilibrium, see Eq. (\ref{new}).  We develop  the necessary conditions to achieve these in this subsection.

{\bf 1. Constraint on the mixed quartic Higgs coupling:}
First, the quartic coupling $\lambda_{HH'}$ of Eq. (\ref{new}) cannot be of order unity.  In its presence, scattering processes such as $H^0 + H^{0*} \rightarrow H^{\prime 0} + H^{\prime 0 *}$ would occur which can bring the two sectors into equilibrium.  The cross section for this process is given by
\begin{equation}
  \sigma(H^0 + H^{0*} \rightarrow H^{\prime 0} + H^{\prime 0 *}) = \frac{|\lambda_{HH'}|^2}{16 \pi s}\sqrt{1-\frac{4 M_H^2}{s}}~.
\end{equation}
The reaction rate can be estimated by multiplying this cross section with the equilibrium number density of $H^0$, a boson, given by
\begin{equation}
    n_b = \frac{\zeta(3)}{\pi^2}g T^3 = 0.243 T^3
\end{equation}
where $g=2$ is the internal degrees of this complex field. Demanding that $\sigma n_b$ remains smaller than the Hubble expansion rate to temperatures down to the Higgs boson mass, we obtain (for $g^* = 112$) the condition
\begin{equation}
    |\lambda_{HH'}| \leq 1.1 \times 10^{-6}~.
    \label{HHp}
\end{equation}
Such a small cross coupling is technically natural, since other interactions do not induce this coupling through quantum corrections. In what follows we shall assume that this condition is always satisfied.

{\bf 2. Constraints on the inflaton coupling to the two sectors:} The inflaton field $\eta$ couples to both sectors, and thus can potentially bring the two sectors into thermal equilibrium.  The cross section for $H^0 + H^{0*} \rightarrow H^{\prime 0} + H^{\prime 0 *}$ mediated by the $\eta$ field (after integrating it out) is given by:
\begin{equation}
     \sigma(H^0 + H^{0*} \rightarrow H^{\prime 0} + H^{\prime 0 *})_\eta = \frac{|4 \lambda_{\eta H}^2 v_\eta^2 - \mu_\eta^2|^2}{M_\eta^4} \frac{1}{16 \pi s}\sqrt{1-\frac{4 M_H^2}{s}}~.
\end{equation}
Choosing parameters such that $T'_{RH}/T_{RH} = 1/3$, we find the condition for this process to be not in equilibrium at $T = M_H$ to be
\begin{equation}
    \left|\frac{\mu_\eta}{M_\eta}\right| \leq 5.4 \times 10^{-4}~.
\end{equation}
(For a more refined treatment of this process and reheating in general, see Ref. \cite{CS}.)
This condition is satisfied easily, since for a low reheat temperature $|\mu_\eta|^2 \ll M_\eta^2$ is anyway necessary.

{\bf 3. Constraint on photon--mirror photon kinetic mixng:} A third constraint arises from the kinetic mixing term in Eq. (\ref{new}) between the two hypercharge gauge fields parametrized by the coupling $\chi$.  This term will result in a kinetic mixing between the photon and the mirror photon with a Lagrangian given by
\begin{equation}
    {\cal L} \supset \frac{\epsilon}{2} F_{\mu\nu}F^{\prime \mu \nu}
\end{equation}
where $\epsilon = \chi \cos^2\theta_W$ with $\theta_W$ being the weak mixing angle.  A shift in the gauge boson fields would remove this mixed kinetic term from the Lagrangian, but in the process mirror fermions would acquire milli-charges under usual electromagnetism \cite{Holdom}.  For example, mirror electron will acquire a coupling to the photon given by $(\epsilon e)(\overline{e'}\gamma_\mu e' A^\mu)$.  Consequently,   scattering processes such as $e^++ e^- \rightarrow e^{\prime +} e^{\prime-}$ would occur with a cross section given in the relativistic limit by
\begin{equation}
    \sigma(e^+ e^- \rightarrow e^{\prime +} e^{\prime-}) = \frac{4 \pi \alpha^2}{3 s}\epsilon^2~.
\end{equation}
These reactions should be out of thermal equilibrium down to temperatures of order the electron mass. 
The number density of fermions in a relativistic plasma is given by 
\begin{equation}
    n_f = \frac{3}{4}\frac{\zeta(3)}{\pi^2} g ^3 = 0.183 T^3
\end{equation}
where $g=2$ is used for the two spin degrees of the fermion. Demanding $\sigma n_f$ to be smaller than the Hubble rate at $T = m_e$ we obtain
\begin{equation}
    |\epsilon| \leq 1.5 \times 10^{-8}~.
    \label{kinlt}
\end{equation}

This constraint on photon-- mirror photon kinetic mixing can be satisfied in the minimal framework, as the particle content of the model is such that no kinetic mixing is induced at lower loop levels.  However, in more extended models this can provide a strong constraint. For example, in a theory where there is a complex scalar field charged under both $Y$ and $Y'$, a nonzero $\chi$ (and therefore $\epsilon$) would be induced given via the renormalization group equation as \cite{BKM}
\begin{equation}
\chi = -\frac{g_Y^2}{48 \pi^2}(Y Y') {\rm ln}\left( \frac{m^2}{\mu^2}\right)
\label{kinetic}
\end{equation}
where $m$ is the mass of the scalar.  This condition would preclude presence of any particle in the theory that is charged under both the SM and SM$'$ gauge symmetries.  This result relies only on the success of asymmetric inflation and is independent of whether  $n-n'$ oscillation is in the observable range or not. 

\section{Modified framework for observable \boldmath${n-n'}$ oscillation}
Here we present a modification of the minimal framework that allows for observable $n-n'$ oscillations which is compatible with asymmetric inflation.  To the particle content of the minimal mirror model discussed in Sec. \ref{sec2} we add a parity even real scalar field $X$ and its parity partner $X'$ which are singlets of the gauge symmetry.  The Lagrangian involving these fields includes the terms
\begin{equation}
    {\cal L}_{\rm new}' = \mu_{\eta X} \,\eta (X^2 - X^{\prime 2}) + \lambda_{\eta X} \eta^2(X^2 + X^{\prime 2}) + \left\{(\overline{N^c}_R Y_XN_R X + \overline{N^{\prime c}}Y_XN'_R X') + h.c.\right\}
    \label{newnew}
\end{equation}
Asymmetric inflation can now be realized in the decays of $\eta \rightarrow XX$ and $\eta \rightarrow X'X'$, and as in the case of $\eta$ decaying into Higgs pairs the ratio of reheat temperatures is given by
\begin{equation}
    \frac{T_{RH}'}{T_{RH}} =  \left|\frac{2 \lambda_{\eta X} v_\eta - \mu_{\eta X}}{2 \lambda_{\eta X} v_\eta + \mu_{\eta X}}\right|~.
\end{equation}
Here we assume that asymmetries arising from decays into Higgs pairs is negligible by choosing $\mu_{\eta H}$ of Eq. (\ref{new}) to be vanishingly small. The advantage here is that there won't be a significant splitting in $M_H^2 - M_{H'}^2$, and thus the two VEVs $v$ and $v'$ can be maintained almost exactly equal.  The mass splitting that arose in the minimal model from the VEV difference given in Eq. (\ref{split}) is therefore absent in the present case.  

The scalar field $X$ is kept in thermal equilibrium with the SM plasma via its decay into a pair of $N$ fields via the coupling $Y_X$ of Eq. (\ref{newnew}), and similarly $X'$ maintains its thermal equilibrium with the SM$'$ plasma through its decay into $N'$.  For this to occur there is a minimum value of the coupling $Y_X$ that is needed. In the approximation $M_N \ll M_X$, the width for the decay $X \rightarrow NN$ is given by
\begin{equation}
    \Gamma(X \rightarrow NN) = \frac{1}{8\pi}{\rm Tr}(Y_X^\dagger Y_X)M_X~.
\end{equation}
Requiring this rate to be of order the Hubble expansion rate at $T \sim M_X$ one obtains (with $g^*(M_X) \approx 112$ and assuming a single Yukawa coupling to be dominant in the decay)
\begin{equation}
    |Y_X| \geq 1.9 \times 10^{-7} \left(\frac{M_X}{\rm TeV}\right)^{1/2}~.
    \label{YX}
\end{equation}

\begin{figure}
    \centering
    \includegraphics[scale=0.7]{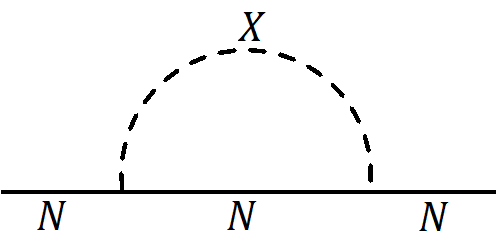}
    \caption{Asymmetric wave function correction for $N$ and $N'$ in the modified model.}
    \label{fig:fig2}
\end{figure}

The constraint of Eq. (\ref{YX}) can be used to estimate the minimum splitting in the mass of $n$ and $n'$.  A one-loop diagram shown in Fig. \ref{fig:fig2} would renormalize the wave function of $N$ differently compared to that of $N'$, since the masses of $X$ and $X'$ scalars differ by order one.  Once this wave-function correction is inserted in the diagram of Fig. \ref{fig:fig1}, the strong coupling $\alpha_s$ and its mirror counterpart $\alpha_s'$ would evolve differently, with the effective $\beta$ function difference given by (see Eq. (\ref{beta}))
\begin{equation}
    \beta_0 - \beta_0' \approx \frac{g_s^2 Y_t^2 Y_D^2Y_X^2}{(16 \pi^2)^3} {\rm ln}\left(\frac{M^2_{X^\prime }}{M_X^2} \right)~.
    \label{beta1}
\end{equation}
Note that this shift arises in the running of the couplings in the momentum range $M_{X'} \leq \mu \leq M_X$. 
The resulting shift in the QCD scale parameters is  $(\Lambda_{QCD} - \Lambda'_{QCD})/\Lambda_{QCD} \approx 10^{-31}$, leading to $m_n-m_{n'} \approx 10^{-31}$ GeV, which is fully consistent with observable $n-n'$ oscillations.

\subsection{Baryon number violating interactions}
For $n-n'$ oscillation to occur, there must be baryon number violation in the theory with an effective quark level operator of the form $(uddu'd'd')$, suppressed by fifth power of an effective mass scale $\Lambda$.  This operator, which breaks both $B$ and $B'$ (baryon number of the mirror world) by one unit preserves $B-B'$ as a global symmetry.  If this symmetry remains unbroken in the full Lagrangian, $n-n'$ oscillation would be permitted, while $n-\overline{n}$ oscillation which breaks $B$ by two units would be forbidden.  Since the energy scale probed by present limits on $n-\overline{n}$ oscillation ($\tau_{n\overline{n}} \geq 2.7 \times 10^8$ sec)  is significantly larger than that probed by $n-n'$ oscillations ($\tau_{nn'} \geq 400$ sec.), for the latter process to be in the observable range it would be desirable to maintain $B-B'$ symmetry to a good approximation.

Before presenting possible UV completions, we write down effective operators with definite chiral and Lorentz structures as follows:
\begin{equation}
    {\cal L}_{\rm eff} = \frac{1}{\Lambda^5} (u_L^i d_L^j)(d_R^k d_R^{\prime \gamma})(u_L^{\prime \alpha}d_L^{\prime \beta}) \epsilon_{ijk} \epsilon_{\alpha \beta \gamma} + h.c.
    \label{d9}
\end{equation}
along with an analogous operator (with a different strength) where the left-handed fermion fields  are replaced by right-handed fields.  It will turn out that these are the two operators that are induced in simple UV complete models presented in the next subsection.  In Eq. (\ref{d9}) a charge conjugation matrix which is not shown  contracts each spinor pair. This effective operator has to be converted to neutron and mirror neutron operators in order to discuss $n-n'$ oscillations.  The hadronic matrix elements relevant for this conversion are defined as
\begin{eqnarray}
\alpha\, u_{nL}(\vec{k}) &=& \epsilon_{ijk} \langle 0|(d_R^i u_R^j)d_L^k|p,\vec{k}\rangle,~~~~~\beta\, u_{nL}(\vec{k}) = \epsilon_{ijk} \langle 0|(d_L^i u_L^j)d_L^k|p,\vec{k}\rangle \nonumber \\
-\alpha\, u_{nR}(\vec{k}) &=& \epsilon_{ijk} \langle 0|(d_L^i u_L^j)d_R^k|p,\vec{k}\rangle,~~-\beta\, u_{nR}(\vec{k}) = \epsilon_{ijk} \langle 0|(d_R^i u_R^j)d_R^k|p,\vec{k}\rangle~.
\label{matrix}
\end{eqnarray}
Here $u_{nL}$ and $u_{nR}$ stand for the left-handed and right-handed neutron spinors respectively. Analogous matrix elements involving the mirror quarks and mirror neutron can be defined, with the coefficients $\alpha$ and $\beta$ identical to those in Eq. (\ref{matrix}) due to mirror parity. These matrix elements have been computed on the lattice rather accurately in the context of nucleon decay. Ref. \cite{Aoki} quotes the continuum limit values of these coefficients to be $\alpha = -0.01257(111)$ GeV$^3$ and $\beta = 0.01269(107)$ GeV$^3$.  (For lattice evaluation of $n-\overline{n}$ hadronic matrix element see Ref. \cite{Wagman}.) The effective Lagrangian of Eq. (\ref{d9}) can then be converted to nucleon level Lagrangian as
\begin{equation}
    {\cal L}_{\rm eff} = \frac{\alpha^2}{\Lambda^5}\left(n \left(\frac{1+\gamma_5}{2}\right) n'\right)~.
    \label{effn}
\end{equation}
Consequently the off-diagonal entry $\delta$ in the Hamiltonian matrix relevant for $n-n'$ oscillation (see Eq. (\ref{Ham})) is $\delta = \alpha^2/(2 \Lambda^5)$, where we have used the fact that in the non-relativistic limit the $\gamma_5$ term does not contribute.  The $n-n'$ oscillation lifetime follows from Eq. (\ref{effn}) as $\tau_{nn'} = \hbar/\delta$ and is given (with the lattice value of $\alpha$ quoted above) by
\begin{equation}
    \tau_{nn'} = 438 ~{\rm sec.} \left(\frac{\Lambda}{35~{\rm TeV}}\right)^5~.
\end{equation}
It is clear that the near future sensitivity of $n-n'$ oscillation would be for the range $\Lambda = (30 - 50)$ TeV, which is what we shall focus on in what follows. 

\subsection{The need for low reheat temperature}

It turns out that the reheat temperature after inflation should be relatively low in the mirror world setup with observable $n-n'$ oscillations.  Recall that the success of asymmetric inflation relies on the SM and SM$'$ sectors remaining thermally decoupled.  The two sectors could be brought into thermal equilibrium through the  effective interaction of Eq. (\ref{d9}) which is needed to induce $n-n'$ oscillation.  Here we derive an upper limit on the reheat temperature $T_{RH}$ arising from demanding that such equilibration does not occur.

In presence of the Lagrangian of Eq. (\ref{d9}) the following two-body to four-body scattering process can occur: 
\begin{equation}
u_L(k_1) + d_L (k_2) \rightarrow \overline{u'_L}(p_1) + \overline{d'_L} (p_2) + \overline{d_R} (p_3) + \overline{d'_R} (p_4)~.
\end{equation}
For a fixed color configuration both in the initial state and in the final state, the spin-averaged squared amplitude for this process, in the limit of massless quarks, is given by
\begin{equation}
    \frac{1}{4}\sum_{{\rm spin}} |M|^2 = \frac{s\, Q_1^2 \,Q_2^2}{4 \Lambda^{10}}
\end{equation}
where $s=(k_1+k_2)^2$, $Q_1^2 = (p_1+p_2)^2$ and $Q_2^2 = (p_3+p_4)^2$. Following the four-body phase space variables defined in Ref. \cite{Weiler} we obtain an analytic expression for the integrated cross section for this process:
\begin{equation}
    \sigma = \frac{1}{(2\pi)^5} \frac{1}{1440} \frac{s^4}{\Lambda^{10}}~.
    \label{phase}
\end{equation}
At temperatures below the effective mass $\Lambda$ and above 1 GeV, this reaction could potentially be faster than the Hubble expansion rate $H(T) = 1.66 \sqrt{g^*(T)}/M_P$.  These two rates become equal at $T = T_*$ where $T_*$ is given by
\begin{equation}
    T_* = 1.44 ~{\rm TeV} \left[\frac{\Lambda} {30~{\rm TeV}} \right]^{10/9}
\end{equation}
which is obtained by equating $\sigma n_f$ to $H(T)$.  Here we used $s=4 \,E^2 = 4\,(3.15T)^2$ (where $E$ is the CM energy of the incident particles) and $g^*(T) \approx 112$. 
Since observability of $n-n'$ oscillations suggests a range $\Lambda = (30-50)$ TeV, for some temperature $T= T_*$ will be realized unless this $T_*$ is above the reheat temperature.  We thus conclude that $T_{RH}$ should obey the condition
\begin{equation}
    T_{RH} \leq 2.54~{\rm TeV} \left[\frac{\Lambda} {50~{\rm TeV}}\right]^{10/9}~.
\end{equation}
This is an improved estimate, including the phase space suppression factors shown in Eq. (\ref{phase}), compared to Ref. \cite{nnprime} and \cite{nasri}. It is interesting to note that the reheat temperature, while low, can still be above the electroweak symmetry breaking scale, which allows the framework to realize baryogenesis via leptogenesis as well as electroweak baryogenesis. 

\subsection{UV completion of the effective operator for \boldmath{$n-n'$} oscillation}
\label{sec3.3}

\begin{figure}
    \centering
    \includegraphics[scale=0.9]{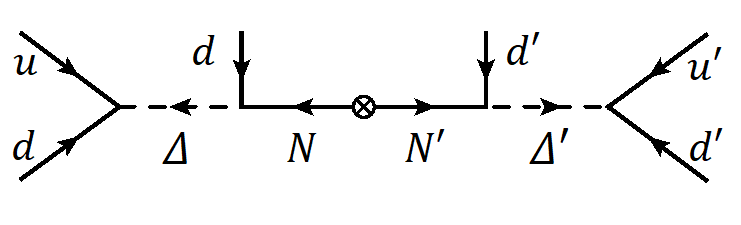}
    \caption{Feynman diagram generating the $n-n'$ oscillation operator in an UV complete theory.}
    \label{fig:fig3}
\end{figure}

We now turn to the generation of ${\cal L}_{\rm eff}$ of Eq. (\ref{d9}) that induces $n-n'$ oscillations from microscopic physics.  When this operator is opened up at tree level, two different topologies arise.  The first one is shown in Fig. \ref{fig:fig3}, which we shall see is essentially the only one that is compatible with asymmetric inflation. This is the diagram suggested in Ref. \cite{nnprime}. The renormalizable Lagrangian generating Fig. \ref{fig:fig3} is
\begin{eqnarray}
{\cal L}_{nn'} &=&  Y_1 \left\{(N_R d_R)\Delta^* + (N'_R d'_R)\Delta^{\prime *}\right)\} + Y_2\left\{(u_L d_L) \Delta + (u'_L d'_L)\Delta'\right\} \nonumber \\
&+& Y_3\left\{(u_R d_R) \Delta + (u'_R d'_R)\Delta'\right\} + M_N (N_R N'_R) + h.c. -M_\Delta^2(|\Delta|^2 + |\Delta^{\prime}|^2)
\label{lag1}
\end{eqnarray}
Here all primed fields transform under the SM$'$ gauge symmetry, while the un-primed fields transform under SM symmetry in identical fashion.  $\Delta$ is a complex scalar with $SU(3)_c \times SU(2)_L \times U(1)_Y$ quantum numbers of $(3,1,-1/3)$, and $\Delta'$ is its mirror partner.  $N_R$ is a fermionic singlet of the SM, with $N_R'$ being its mirror partner.  Note that the Lagrangian of Eq. (\ref{lag1}) breaks baryon number $B$ and mirror baryon number $B'$, but it conserves $B-B'$ symmetry. This can be seen by assigning $(\Delta,\,N_R)$ fields baryon numbers of $(-2/3,\,-1)$, along with $B'$ charges of $(-2/3,\, -1)$ for the $(\Delta',\, N_R')$ fields.  If the $B-B'$ symmetry is maintained, $n-\overline{n}$ oscillations would be forbidden, but $n-n'$ oscillations would be allowed. This symmetry may be broken down to a $Z_2$ subgroup by adding Majorana masses for the $N_R$ and $N_R'$ fields of the type 
\begin{equation}
{\cal L}_{\rm Maj} =
\delta M_N(N_R N_R + N'_R N'_R) + h.c.
\label{Maj}
\end{equation}
In this case the mass parameter $\delta M_N$ will have to be of order $10^{-5} M_N$ or smaller
for observable $n-n'$ oscillations, since in the presence of Eq. (\ref{Maj}) $n-\overline{n}$ oscillations would occur which is constrained by experiments ($\tau_{n-\overline{n}} \geq 2.8 \times 10^8$ sec.) at a level  which is about five orders of magnitude stronger than $n-n'$ oscillations ($\tau_{nn'} \geq 400$ sec.). We shall entertain including such a small $\delta M_N$ term in the next section, but focus here on the exact $B-B'$ symmetric limit where the RHS of Eq. (\ref{Maj}) is identically zero.

Integrating out the $\Delta$ and $\Delta'$ fields in Eq. (\ref{lag1}) one would obtain an effective four-Fermi Lagrangian carrying nonzero baryon number given by
\begin{eqnarray}
 {\cal L}_{d=6} &=& \frac{Y_1Y_2}{M_\Delta^2}\left\{(u_L^i d_L^j)(d_R^k N_R)\epsilon_{ijk} + (u_L^{\prime \alpha} d_L^{\prime\beta})(d_R^{\prime\gamma} N'_R)\epsilon_{\alpha \beta \gamma} \right\} \nonumber \\
 &+& \frac{Y_1Y_3}{M_\Delta^2}\left\{(u_R^i d_R^j)(d_R^k N_R)\epsilon_{ijk} + (u_R^{\prime \alpha} d_R^{\prime\beta})(d_R^{\prime\gamma} N'_R)\epsilon_{\alpha \beta \gamma} \right\} + M_N (N_R N_R') + h.c.
 \label{lag2}
\end{eqnarray}
Here we have not displayed $B$ conserving terms, which are irrelevant for $n-n'$ oscillations. If we also integrate out the $N_R$ and $N'_R$ fields from here, one would obtain the effective six-fermion Lagrangian of Eq. (\ref{d9}), with the identification \begin{equation}
    \Lambda^5 = \frac{M_\Delta^4 M_N}{Y_1^2Y_2^2}~
    \label{Lambda}
\end{equation}
along with a similar term where all quarks are right-handed and the coupling $Y_2$ is replaced by $Y_3$.  The mass scale $\Lambda$ should be in the range $(30-50)$ TeV for observable $n-n'$ oscillations, which sets upper limits on the masses of $\Delta$ and $N$.  As an example, consider $Y_1 = Y_2 = 0.1$, $M_\Delta = 20$ TeV and $M_N = 30$ GeV, which would lead to $\Lambda = 34$ TeV, which is consistent with all experimental constraints.  Of course, other choices of the Yukawa couplings $Y_{1,2}$ are possible, leading to different masses of $\Delta$ and $N$ fields. Lower values of the Yukawa couplings would lower the mass of $\Delta$, in which case it may be within reach of the LHC.  However, as we shall see in the next subsection, the mass of $N$ has to lie in the range $(1-100)$ GeV, or else spin-flip transitions involving the $N$ field would bring the SM sector and the mirror sector into equilibrium.

\begin{figure}
    \centering
    \includegraphics[scale=0.8]{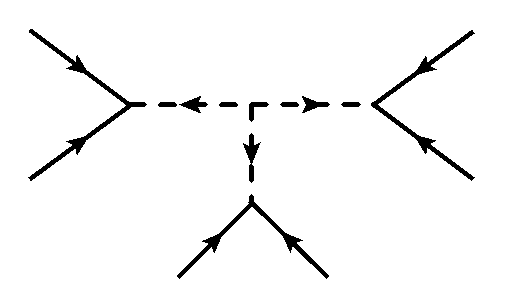}
    \caption{A second possible topology for inducing the $d=9$ operator for $n-n'$ oscillation.}
    \label{fig:fig4}
\end{figure}
A second possible way of generating the six-fermion operator of Eq. (\ref{d9}) is by integrating out only scalar fields, as shown in Fig. \ref{fig:fig4}.  However, in this topology, at least one of the scalars should carry hypercharge under both SM and SM$'$. For example, the three scalars could be $(\Delta_{ud},\, \Delta_{u'd'},\, \Delta_{dd'})$ having Yukawa couplings to quarks of the type  $\Delta_{ud} (ud) + \Delta_{u'd'}(u'd') + \Delta_{dd'}(dd')$. The field $\Delta_{dd'}$ carries both hypercharges $Y$ and $Y'$ in this case. This would lead to an induced kinetic mixing term between the photon and mirror photon at the level of $10^{-3}$, see Eq. (\ref{kinetic}), which would violate the bound derived in Eq. (\ref{kinlt}). The same remark applies to the general topology of the type shown in Fig. 3 with fermions and scalars, unless the fermion is a singlet of both gauge symmetries.  Topologies of Fig. \ref{fig:fig3} and \ref{fig:fig4} are the only ones that allow for inducing the six-fermion operator of Eq. (\ref{d9}) at tree level. We thus conclude that the diagram of Fig. \ref{fig:fig3} is the only way to generate $n-n'$ oscillations at the observable level in the framework of asymmetric inflation.

\subsection{Spin-flip transition constraint}

The SM and SM$'$ sectors could be brought into thermal equilibrium via spin flip transition involving the $N$ and $N'$ fields.  As shown in Sec. \ref{sec3.3}, $N_R$ and $(N'_R)^c$ form right-handed and left-handed components of a Dirac fermion. Since $N$ is in equilibrium with the SM plasma and $N'$ in equilibrium with the SM$'$ plasma, any interaction that leads to spin-flip of this Dirac fermion could bring the two sectors into equilibrium.  In order to derive the constraint for this not to happen we focus on the first term of the $d=6$ four-Fermi Lagrangian of Eq. (\ref{lag2}).  This term can be recast after a Fierz rearrangement as
\begin{equation}
    {\cal L}_{d=6} = -\frac{Y_1Y_2}{8 M_\Delta^2}\left(\overline{N^c}\gamma_\mu(1-\gamma_5)d^j \right)\left(\overline{u^c} \gamma^\mu(1+\gamma_5)d^k \right)\epsilon_{ijk} + ...
\end{equation}
We have computed the spin-flip cross section for the process $d(p_1) + d(p_2) \rightarrow u^c(p_3) + N(p_4,\lambda)$ (with $\lambda$ denoting the helicity eigenvalue) following the formalism developed in Ref. \cite{gandhi} and find this cross section to be
\begin{equation}
    \sigma({\rm spin~filp}) = \frac{M_N^2}{ 192 \pi \tilde{\Lambda}^4}\left(1-\frac{M_N^2} {s} \right)^2~.
\end{equation}
Here we have defined $\tilde{\Lambda}^2 = M_\Delta^2/(Y_1Y_2)$ obeying the relation $\Lambda^5 = \tilde{\Lambda}^4 M_N$ (with $\Lambda$ defined in Eq. (\ref{d9})). We demand that $\sigma \times n_f$ be less than the Hubble rate $H(T) = 1.66 \sqrt{g^*(T)}/M_P$ with $n_f = 0.118 T^3$. The two rates become equal if, for example, we choose $\tilde{\Lambda} = 100$ TeV and $M_N = 100$ GeV (corresponding to $\Lambda = 25$ eV) at $T = 56$ GeV, with the spin-flip rate exceeding the Hubble rate at higher temperatures.  This choice would require the reheat temperature to be less than 56 GeV so that this cross-equilibrium is not established.  As the mass of $N$ field is lowered, the reheat temperature can be raised.  If we choose $M_N = 30$ GeV and $\tilde{\Lambda} = 150$ TeV (so that $\Lambda = 27$ TeV), the equality of the two rates occurs at $T = 2.7$ TeV, which is consistent with observable $n-n'$ oscillation and asymmetric inflation.  The scenario prefers lower mass of the $N$ field, in the range of $(1-100)$ GeV within the framework. (The lower limit arises from nucleon stability, as proton would decay into $N$ and a $\pi^+$ for lower masses of $N$.) Since the exchange of such a singlet fermion field is essentially unique in inducing $n-n'$ oscillation at an observable level, the existence of a light $N$ in this mass range may be regarded as a prediction of the framework. This opens up the possibility that this singlet fermion can be potentially discovered at the LHC.

\subsection{\boldmath$n-n'$ mass splitting in the UV complete theory}

\begin{figure}[h]
    \centering
    \includegraphics[scale=0.55]{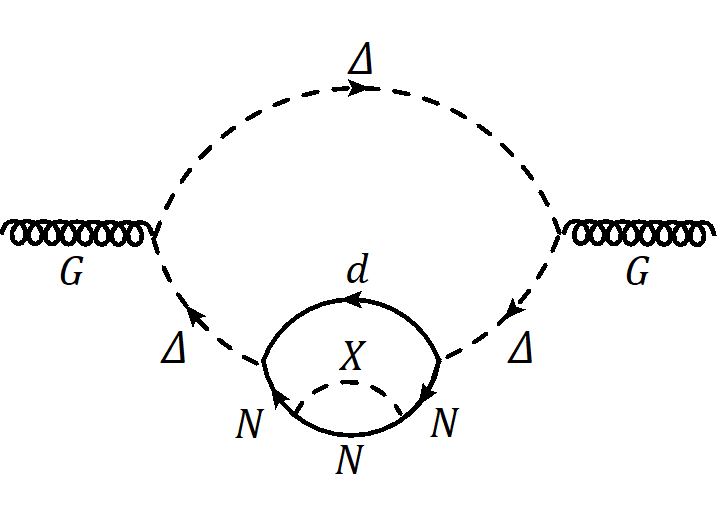}
    \caption{Leading contribution to the $n-n'$ mass splitting in the UV complete theory.}
    \label{fig:fig6}
\end{figure}

The new interactions of Eq. (\ref{lag1}) can potentially induce a mass splitting between the neutron and mirror neutron.  We find the most important diagram in the model that induces such a splitting to be the one given in Fig. \ref{fig:fig6}.  Here the scalar $\Delta$, being colored, couples directly to the gluon, and it also has coupling to $N$ and $d$.  $N$ in turn couples to the $X$ scalar, which has a different mass compared to its $X'$ mirror. Following the discussions around Eq. (\ref{beta}), we now obtain for the difference in the effective QCD and QCD$'$ beta functions
\begin{equation}
    \beta_0 - \beta_0' \approx \frac{g_s^2 Y_X^2 Y_1^2}{(16 \pi^2)^2} {\rm ln}\left(\frac{M^2_{X^\prime }}{M_X^2} \right)~.
    \label{beta3}
\end{equation}
Now, $|Y_X| \geq 1.9 \times 10^{-7}$ is needed for $X$ to be in equilibrium with $N$ for $X$ mass of about 1 TeV, see Eq. (\ref{YX}).  If the coupling $Y_1$ is of order one, the two QCD scales would differ by a factor $10^{-15}$, leading to unacceptably large $n-n'$ mass splitting.  One could take small values of $Y_1$, but in this the scale $\Lambda$ of Eq. (\ref{d9}) that controls $n-n'$ oscillations will become large. A judicious choice which avoids this conflict is to take $Y_1 \approx 10^{-4}$, $M_\Delta = 2$ TeV, $Y_2 = 1$ and $M_N = 10$ GeV, in which case $\Lambda = 28$ TeV (see Eq. (\ref{Lambda})). Our estimate for this choice is $m_n - m_{n'} \approx 10^{-23}$ GeV, which is just about sufficient for unsuppressed oscillations. Note that such a choice of parameters would make the colored scalar relatively light, which may be within reach of the LHC.

It should be pointed out that when the $n-n'$ mass splitting is of order $10^{-18}$ GeV, which happens quite naturally in the model, it may be possible to cancel this energy difference by applying magnetic filed in the search experiment of order 10 Gauss.  This is an interesting way to achieve an MSW-like resonance in the oscillation even when there is small mass splitting. It would remain to be a challenge, since theoretically the mass splitting cannot be computed with accuracy needed for experiments.

There is one simple modification of the model presented here that would make the $n-n'$ mass splitting well below $10^{-24}$ GeV without sacrificing the successful features of the model.  Suppose that there is an additional scalar singlet $Y$ and its mirror partner $Y'$ in the model.  One could now couple $Y$ and $Y'$ asymmetrically to the inflaton field $\eta$ through the couplings
\begin{equation}
    {\cal L}'_{\rm new} = \mu_{\eta Y} \,\eta \,(Y^2 - Y^{\prime 2}) + \lambda_{\eta Y} \eta^2(Y^2 + Y^{\prime 2}) + \lambda_{XY}(X^2 Y^2 + X^{\prime 2}Y^{\prime 2})~.
\end{equation}
If the quartic coupling $\lambda_{XY}$ is of order $10^{-6}$ or larger, that would keep $X$ and $Y$ in equilibrium (and similarly in the mirror sector).  It is the masses of $Y$ and $Y'$ that are split by order one in this case, and not that of $X$ and $X'$.  Note that $Y$ does not have a significant Yukawa coupling with $N$ in this modification.  The diagram of Fig. \ref{fig:fig6} will not lead to an asymmetric running of $\alpha_s$ and $\alpha_s'$.  One would need to insert a $Y$ loop connected to the $X$ line of Fig. \ref{fig:fig6} to generate an asymmetry. However, this four-loop diagram would be suppressed by an additional factor of $\lambda_{XY}/(16 \pi^2)$ compared to Eq. (\ref{beta3}), which could bring in a suppression of order $10^{-8}$. In this case the Yukawa coupling $Y_1$ that appears in Eq. (\ref{beta3}) can be of order 0.1, and the parameter space for $n-n'$ oscillation would open up considerably.
    
\section{A connection with neutrino mass generation}

Since the almost unique way of generating the $d=9$ operator of Eq. (\ref{d9}) that induces $n-n'$ oscillation is by the exchange of a neutral singlet fermion $N$, it is very tempting to explore if this field can be identified as the right-handed neutrino.  If it has the canonical couplings of the seesaw mechanism, as shown in Eq. (\ref{seesaw}), this identification is problematic, since after electroweak symmetry breaking $N$ mixes with the light neutrino.  The effective $d=6$ operator $uddN$ would lead to proton decay $p \rightarrow \overline{\nu} \pi^+$, the constraint from which would require the scale $\Lambda$ to be close to the GUT scale.  One possible way out is to identify $N$ appearing in the $uddN$ operator as one of the three right-handed neutrinos which decouples from the seesaw mechanism \cite{babu}.  Realistic neutrino masses and mixings can be produced involving only two right-handed neutrinos.  In this scenario, one could adopt resonant leptogenesis to explain the observed baryon asymmetry of the universe.

\begin{figure}[h]
    \centering
    \includegraphics[scale=0.65]{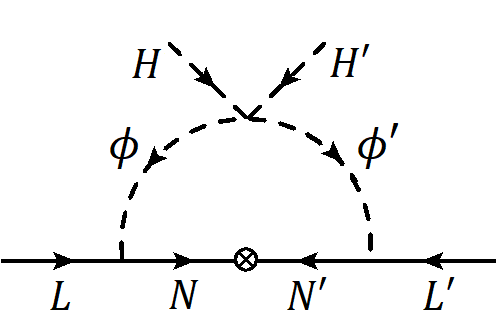}
    \caption{One-loop diagram inducing Dirac neutrino masses. Here $\phi$ is a second Higgs doublet that carries baryon number.}
    \label{fig:fig6}
\end{figure}

A second alternative, which we develop here further, would treat all three right-handed neutrino fields on the same footing, but their Dirac Yukawa couplings involve a second Higgs doublet $\phi$ (and its mirror partner $\phi')$ that does not acquire a VEV.  The new Yukawa couplings are
\begin{equation}
    {\cal L}'_{N} =  \left(\overline{L}Y_D \tilde{\phi} N_R + \overline{L'}Y_D \tilde{\phi'} N'_R \right) + h.c.
    \label{Dirac}
\end{equation}
In this case $(B-B')$ can be maintained as a good symmetry, as can be seen by assigning $B$ of $+1$ to $\phi$ (and similarly $B'$ of $+1$ to $\phi'$) field. The Higgs potential would now contain a $(B-B')$ conserving quartic coupling given as
\begin{equation}
V \supset \lambda_{H\phi}(H^\dagger \phi)(H^{\prime \dagger}\phi') + h.c.
\end{equation}
Although such a coupling could potentially bring the SM and SM$'$ into equilibrium, this can be avoided by taking the (common) masses of $\phi$ and $\phi'$ above the reheat temperature of order TeV.  The neutrinos will now acquire Dirac masses with the right-handed partners being the $\nu'$ fields of the mirror sector.  These arise through the one-loop diagram shown in Fig. \ref{fig:fig6}. We estimate the induced Dirac masses to be
\begin{equation}
    m_\nu \simeq \frac{Y_D^2}{16 \pi^2}\frac{\lambda_{H\phi} v^2}{M_\phi^2}M_N~.
\end{equation}
As an example of consistent parameter set, let $M_N = 10$ GeV, $\lambda_{H\phi} = 10^{-3}$, $Y_D = 1$ and $M_\phi = 10^5$ GeV, which would lead to $m_\nu \sim 0.1$ eV.  Note that the quartic coupling $\lambda_{H\phi}$ would induce a term of the type $(H^\dagger H)(H^{\prime \dagger} H')$ through a one-loop diagram with a strength  $\lambda_{HH'} \sim \lambda_{H\phi}^2/(16\pi^2)$, which for the choice of parameters quoted above is of order $10^{-8}$. This is below the critical value derived in Eq. (\ref{HHp}) that would bring the two sectors into equilibrium.

\begin{figure}[h]
    \centering
    \includegraphics[scale=0.65]{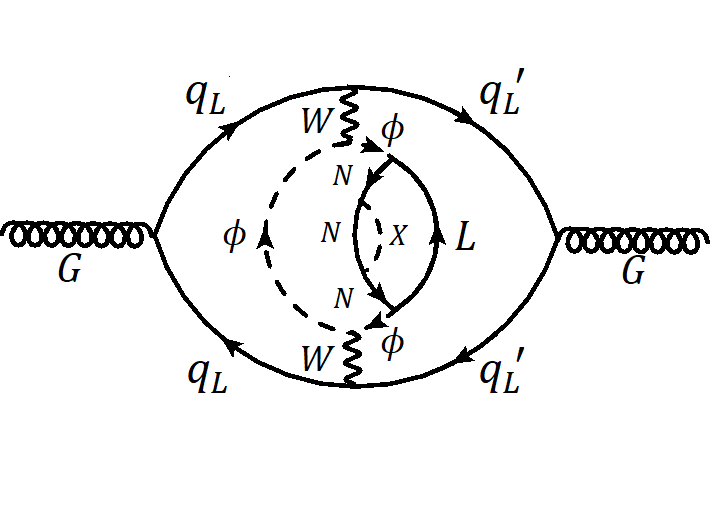}
    \caption{A higher loop diagram inducing $n-n'$ mass splitting in the model with a second Higgs doublet $\phi$.}
    \label{fig:fig6}
\end{figure}

In this version, diagrams such as the one shown in Fig. \ref{fig:fig7} would lead to a splitting of the masses of $n$ and $n'$. The shift in the QCD and QCD$'$ beta function coefficients arising from this diagram can be estimated to be
\begin{equation}
    \beta_0 - \beta_0' \approx \frac{g_s^2 g^4 Y_D^2 Y_X^2 }{(16 \pi^2)^4} {\rm ln}\left(\frac{M^2_{X^\prime }}{M_X^2} \right)~.
    \label{beta3}
\end{equation}
Owing to the presence of more loops, the induced mass splitting can be below $10^{-24}$ GeV. For example, the choice $Y_D = 10^{-1}, Y_X = 10^{-7}$ would lead to $m_n-m_{n'} \approx 10^{-24}$ GeV.

\subsection{A scenario with observable \boldmath${n-n'}$ and \boldmath${n-\overline{n}}$ oscillations}

\begin{figure}[h]
    \centering
    \includegraphics[scale=0.65]{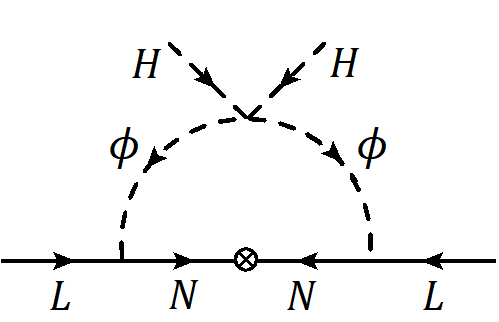}
    \caption{One-loop diagram inducing Majorana neutrino masses as in the scotogenic model.}
    \label{fig:fig7}
\end{figure}

There is an interesting possibility in the model with a second Higgs doublet $\phi$ to realize $n-n'$ oscillation and $n-\overline{n}$ oscillations at the observable level simultaneously.  Consider the addition of the Majorana masses to the $N$ and $N'$ fields as shown in Eq. (\ref{Maj}).  These mass terms $\delta M_N$ should be of order $10^{-5}$ times the leading Dirac mass terms $M_N$.  The $(B-B')$ symmetry is now broken down to a $Z_2$ subgroup. For this choice of parameters both $n-n'$ and $n-\overline{n}$ oscillations can be near the current experimental limits. The Higgs potential would now have an additional quartic coupling given by
\begin{equation}
    V \supset \lambda'_{H\phi}\left\{(H^\dagger \phi)^2 +(H^{\prime \dagger} \phi')^2\right\} + h.c.
\end{equation}
This term  preserves a $Z_2$ subgroup of $B-B'$ symmetry, which guarantees the proton stability. Note that the lightest of the $\phi$ field is not a dark matter candidate, as it can decay into $N$ and a neutrino. The $N$ field is unstable as it decays into three quarks. Mirror baryons are the only dark matter candidates in this setup. In addition to the diagram of Fig. \ref{fig:fig6} that induces Dirac masses for the neutrinos, now there would be also Majorana masses as in the scotogenic model \cite{Ma}. The use of scotogenic model to avoid the proton decay problem in the case of $n-\bar{n}$ oscillation was discussed in ~\cite{bhupal}. The relevant diagram is shown in Fig. \ref{fig:fig7}. This Majorana mass can be estimate to be
\begin{equation}
    m_\nu^{\rm Maj} \simeq \frac{Y_D^2}{16\pi^2} \frac{\lambda'_{H\phi}v^2}{M_\phi^2}\delta M_N~.
\end{equation}
These induced Majorana masses are much smaller than the Dirac masses given in Eq. (\ref{Dirac}), since $\delta M_N \approx 10^{-5} M_N$, and since we can take $\lambda'_{H\phi} \approx 10^{-5} \lambda_{H\phi}$ quite naturally.  Although the neutrino is now a pseudo-Dirac particle, the mass-splitting between the two states could be of order $10^{-11}$ GeV, well below experimental requirements.

\section{Summary and conclusion} 

We have examined in this paper theoretical constraints arising from observable $n-n'$ oscillation signals and asymmetric inflation.  One of the main constraints arises from the mass splitting induced by parity violation needed for asymmetric inflation in the masses of $n$ and $n'$.  We have found that the post-inflationary reheat temperature has an upper bound of about 2.5 TeV, and that a neutral singlet fermion should exist with a mass below 100 GeV for observable $n-n'$ oscillation strength.  

We have also tried to connect the neutral fermion present in the theory to neutrino mass generation. Two scenarios are realized: one where the neutrinos are Majorana particles and the singlet fermion inducing $n-n'$ oscillations is decoupled from the other two right-handed neutrinos, and another where the neutrinos are Dirac particles with their masses arising radiatively from one-loop diagrams. 
This framework suggests the possibility of observing both $n-n'$ and $n-\overline{n}$ oscillations in near future experiments.

\section*{Acknowledgement}  We thank Yuri Kamyshkov for discussions. The work of KSB is supported by the US Department of Energy Grant No. DE-SC 0016013  and that of RNM is supported by NSF grant No. PHY-1914631.

\end{document}